\documentclass[12pt]{article}

\usepackage{epsfig,amsmath,amssymb,latexsym}

\providecommand{\beqa}{\begin{eqnarray}}
\providecommand{\eeqa}{\end{eqnarray}}
\providecommand{\we}{\wedge}
\providecommand{\tr}{\text{tr}}
\providecommand{\MT}{M$5_2$ }

\providecommand{\SF}{$\mathbf{\Sigma}_4$}
\providecommand{\STN}{$\mathbf{\Sigma}_2^{0}$}

\def\Orb{{\mathbf{S}^1/\mathbf{Z}_2}}
\def\Z2{{\mathbf{Z}_2}}

\def\cF{{\cal{F}}}
\def\cM{{\cal{M}}}

\def\cS{{\cal{S}}}

\numberwithin{equation}{section}

\begin{document}

\thispagestyle{empty}
\rightline{MIFP-05-28}
\rightline{hep-th/0510066}

\begin{center}
{\bf \LARGE Heterotic Cosmic Strings}
\end{center}

\vspace{1.3truecm} \centerline{Katrin Becker$^{a,b,c,}
$\footnote{kbecker@radcliffe.edu}, Melanie
Becker$^{a,b,c,}$\footnote{mbecker@radcliffe.edu}, Axel
Krause$^{b,c,}$\footnote{akrause@fas.harvard.edu}}

\vspace{.7truecm}

\centerline{{\em $^a$Radcliffe Institute, Harvard University,
Cambridge, MA 02138, USA}}

\vspace{.2truecm}

\centerline{{\em $^b$Jefferson Physical Laboratory, Harvard
University, Cambridge, MA 02138, USA}}

\vspace{.2truecm}

\centerline{{\em $^c$George P.~\& Cynthia W.~Mitchell Institute for
Fundamental Physics,}}
\centerline{{\em Texas A\&M University, College Station, TX 77843,
USA}}

\vspace{1.0truecm}

%%%%%%%%%%%%%%%%%%%%%%%%%%%%%%%%%%%%%%%%%%%%%%%%%%%%%%%%%%%%%%%%%%

\begin{abstract}
We show that all three conditions for the cosmological relevance of
heterotic cosmic strings, the right tension, stability and a
production mechanism at the end of inflation, can be met in the
strongly coupled M-theory regime. Whereas cosmic strings generated
from weakly coupled heterotic strings have the well known problems
posed by Witten in 1985, we show that strings arising from M5-branes wrapped around 4-cycles (divisors) of a Calabi-Yau in heterotic M-theory compactifications, solve these problems in an elegant fashion.
\end{abstract}

\noindent
%PACS: 04.65.+e, 11.25.Mj, 11.25.Yb (Noch zu Aendern)\\

\newpage
\pagenumbering{arabic}

\section{Introduction}

It has been known for a long time that COBE data require the
effective or fundamental tension $\mu$ of a cosmic string to be
given by $G_N\mu \simeq 10^{-6}$ if the scaling solution of the
cosmic string network is assumed to be the prime source for
density perturbations which seed galaxy formation. The option that
cosmic strings are primarily responsible for structure formation
has, however, been ruled out by more recent CMB data. More
precisely it has been shown \cite{Contaldi:1998qs} that present
CMB data \cite{Lee:2001yp} constrain the contribution of a cosmic
string network to the CMB anisotropies to be less than 20\%. This
leads to a slightly tighter upper bound
\beqa
G_N\mu \lesssim 2\times 10^{-7},
\label{Bound}
\eeqa
on the cosmic string tension. The bound can equivalently be
written as $\sqrt{\mu} \lesssim 5.5 \times 10^{15}$\,GeV and
indicates that the energy scale associated with the cosmic string
tension should be roughly of the order of the GUT scale (for
recent reviews on cosmic strings see \cite{Vachaspati:2000cq},
\cite{Polchinski:2004ia}, \cite{Vilenkin:2005jg},
\cite{Davis:2005dd}).

For the weakly coupled heterotic string $\mu$ equals the
fundamental string's tension $T=1/2\pi\alpha'$ which is given by
the string scale squared $M_s^2$. Since $M_s\simeq 10^{18}$\,GeV
we are 2.5 orders of magnitude above the required energy scale and
would hence violate the bound (\ref{Bound}). Another way to see
this is to remember the fact that in the weakly coupled heterotic
string gravitational and gauge couplings are tightly related,
$4\kappa_{10}^2=\alpha' g_{10}^2$ since both originate at the
level of the trilinear interactions of the closed heterotic
string. This same origin also implies that both gravity and the
gauge fields live in the total 10d spacetime (this no longer holds
for the strongly coupled heterotic string) and therefore both
couplings reduce in the same way to the corresponding 4d
couplings. With $\alpha_{GUT}\simeq 1/25$ being the 4d gauge
coupling whose value follows from the unification of all gauge
forces, we obtain
\beqa
G_N \mu = {\alpha' \alpha_{GUT} \over 8} \mu \simeq 8 \times
10^{-4},
\eeqa
which clearly violates the bound (\ref{Bound}). Consequently
weakly coupled heterotic fundamental strings cannot lead to viable
cosmic strings, as has been realized by Witten twenty years ago
\cite{Witten:1985fp}\footnote{Recently the appearance of open
heterotic SO(32) strings has been discovered in
\cite{Polchinski:2005bg}. It would be interesting to understand
their potential role in cosmology.}.

In type II theories the string-scale can be lowered down to the
TeV scale. This allows for a large range of cosmic string tensions
below the GUT scale in compliance with the observational bound
\cite{Sarangi:2002yt}, \cite{Jones:2003da}. However, this large
range for the fundamental string-scale weakens the predictivity of
type II cosmic strings. Their tensions might well be below
observational verification. To have a more predictive framework,
we will now consider the strongly coupled heterotic string where
the Planck-scale is fixed. The fact which makes this theory very
interesting for cosmic strings is that the gravitational coupling
scale which determines the M2 and M5 brane tensions,
\beqa
\kappa_{11}^{2/9}\simeq\frac{1}{2 M_{GUT}} \; ,
\label{GUTScale}
\eeqa coincides roughly with the 4d GUT scale $M_{GUT}\simeq
3\times 10^{16}$\,GeV \cite{Banks:1996ss}. Hence we can expect
that the effective tensions of cosmic strings arising from
suitably wrapped M2 and M5 branes might be close to the bound
(\ref{Bound}). This is our main reason to focus on the strongly
coupled heterotic string or heterotic M-theory for
short\footnote{Another decisive virtue is its realistic
phenomenology which has gained renewed interest recently
\cite{Forste:2004ie}, \cite{Kobayashi:2004ya},
\cite{Donagi:2004ub}, \cite{Braun:2005ux}, \cite{Braun:2005bw},
\cite{Blumenhagen:2005ga}, \cite{Raby:2005vc}.}. We will show in
this paper that all three criteria -- tension, stability,
production at the end of inflation
-- can be satisfied in the M5 brane case.

\section{Cosmic String Candidates from Wrapped M2 and M5 Branes}

Heterotic M-theory contains only two extended objects, the M2 and
the M5 brane which we are exploring as candidates for heterotic
cosmic strings. The theory also contains 10-dimensional boundaries
which might loosely be regarded as M9 branes. They fill, however,
all of the 4-dimensional spacetime and can therefore not generate
cosmic strings. For the generation of gauge cosmic strings, which we are not investigating here, this is another matter as the Yang-Mills vector bundles are localized precisely on the M9's. It should be interesting to explore this question in the future. Generating a cosmic string from wrapped M2 or M5 branes means that these branes must extend along a time-like and a space-like direction, $t,x$, into four-dimensional spacetime.

We consider heterotic M-theory compactified on $X\times\Orb$,
where $X$ is a Calabi-Yau threefold. The resulting flux
compactification geometry has in the simplest case a Calabi-Yau
which is conformally deformed by a warp-factor generated from the
background $G^{(2,2,0)}$ flux \cite{Curio:2000dw},
\cite{Curio:2003ur}, \cite{Krause:2001qf} (see also
\cite{Witten:1996mz}). We will now consider wrapping M2 and M5
branes over suitable cycles in this 7-dimensional
flux-compactification background and start by listing all possible
candidates for obtaining cosmic strings in four dimensions.

Let us begin with those configurations which are considered BPS in
the flat spacetime limit. These are the M2 brane transverse to the
M9's and the M5 brane parallel to them. The M2 brane which
stretches along the $\Orb$ interval produces in the limit of
vanishing orbifold length $L$, i.e.~the weakly coupled limit, a
fundamental heterotic string. Since the fundamental heterotic
string is a closed string, we learn that the M2 brane worldvolume
must have the following topology
\beqa
M2_{\perp}:
\; \underbrace{\mathbf{R}^1\times\mathbf{S}^1}_{\text{cosmic string
loop}} \!\!\!\! \times \,\, \Orb \; ,
\eeqa
giving rise to a cosmic string loop.

The parallel M5 brane needs to wrap a 4-cycle {\SF} on $X$ to
produce a string-like object. For this we need to adopt a Calabi-Yau with non-vanishing $b_4(X) = 2h^{3,1}+h^{2,2} = h^{1,1} \ne 0$ which is the generic case. The topology of the M5 brane worldvolume will then be
\beqa M5_{\parallel}:
\underbrace{\mathbf{R}^1\times\mathbf{R}^1}_{\text{$\infty$-extended
cosmic string}} \!\!\!\!\!\!\!\!\!\!\! \times \,\,
\mathbf{\Sigma}_4 \; ,
\eeqa
where the two non-compact time and space directions are along the
two M5 brane dimensions which extend into the 4-dimensional
spacetime and create naturally an infinitely extended cosmic
string.

One might also contemplate parallel M2 branes by wrapping the M2
not along $\Orb$ but instead on a 1-cycle of $X$. This would also
create a string but can be ruled out because the Calabi-Yau
threefold has vanishing first Betti number, $b_1(X) = 2h^{1,0} =
0$, hence possesses no 1-cycles on which the M2 could be wrapped
(we will not consider non-simply connected Calabi-Yau's). More
interesting are the transverse M5 branes which wrap one of the
$b_3(X) = 2(h^{3,0}+h^{2,1}) = 2(1+h^{2,1}) \ne 0$ 3-cycles
$\mathbf{\Sigma}_3$ and have topology
\beqa
M5_{\perp}:
\underbrace{\mathbf{R}^1\times\mathbf{R}^1}_{\text{$\infty$-extended
cosmic string}} \!\!\!\!\!\!\!\!\!\!\! \times \,\,
\mathbf{\Sigma}_3\times\Orb \; .
\eeqa
The resulting cosmic string would again be an infinitely extended
cosmic string.

We will next derive the tensions of the cosmic string and compare
them with the constraint (\ref{Bound}). An important role will be
played by the warped background which influences the tension. The
observational bound will eliminate the $M2_{\perp}$ candidate and
leave us with the two M5 brane candidates.

\section{Cosmic String Tensions}

\subsection{$\text{M2}_{\perp}$ Brane Case}

Let us begin with the $\text{M2}_\perp$ brane. To determine the
effective tension of the associated cosmic string we take the
Nambu-Goto part of the $\text{M2}_\perp$ brane action
\beqa
S_{M2} = \tau_{M2}\int_{\mathbf{R}^1} \!\!\! dt
\int_{\mathbf{S}^1} \!\!\! dx \int_0^L \!\!\!\! dx^{11}
\sqrt{-\det h_{ab}} + \hdots \; ,
\eeqa
and integrate it over the compact dimension $x^{11}$. Here
$a,b,\hdots = t,x,x^{11}$ and $L$ is the length of the $\Orb$
interval. We adopt a static gauge for the embedding of the
$\text{M2}_\perp$ into 11-dimensional spacetime which gives us for
the induced metric ($I,J=0,\hdots,9,11$)
\beqa
h_{ab} \equiv \frac{\partial X^I}{\partial x^a} \frac{\partial
X^J}{\partial x^b} G_{IJ} = \delta_a^I \delta_b^J G_{IJ} \; .
\eeqa
The 11d metric $G_{IJ}$ is given by the warped $G$-flux
compactification background sourced by the boundary M9's
\cite{Curio:2000dw}, \cite{Curio:2003ur}, \cite{Krause:2001qf}
\beqa
ds_{11}^2 = G_{IJ} dx^I dx^J = e^{-f(x^{11})} g_{\mu\nu} dx^\mu
dx^\nu + e^{f(x^{11})} \Big(g(X)_{lm} dy^l dy^m + dx^{11}
dx^{11}\Big) \; ,
\label{FluxMetric}
\eeqa
where the warp-factor is given by\footnote{The charge $Q_v$ had
been denoted $\cS_v$ in \cite{Curio:2000dw}, \cite{Curio:2003ur},
\cite{Krause:2001qf}.}
\beqa
e^{f(x^{11})} = (1-x^{11}Q_v)^{2/3}
\eeqa
with visible M9 brane charge
\beqa
Q_v = -\frac{1}{8\pi V_v}
\Big( \frac{\kappa_{11}}{4\pi} \Big)^{2/3}
\int_{X_v} J \we (\tr F\we F-\frac{1}{2}\tr R\we R)
\eeqa
which sources the $G^{(2,2,0)}$ flux component. $X_v$ and $V_v$
denote the Calabi-Yau and its volume at the location of the
visible M9, $J$ its K\"ahler-form and $F$ resp.~$R$ the Yang-Mills
and curvature 2-forms, again on the visible M9.

Notice that we are taking the flux background which incorporates
only the backreaction of the M9's but not that of extra \MT branes
in the bulk. The extra \MT branes would wrap genus zero
holomorphic 2-cycles on $X$ and fill all of 4-dimensional
spacetime so shouldn't be confused with the $\text{M5}_\parallel$,
$\text{M5}_{\perp}$ brane candidates for cosmic strings. Though
the backreaction of the \MT branes is known \cite{Curio:2000dw},
\cite{Curio:2003ur}, \cite{Krause:2001qf}, their neglect is
justified when we want to focus on a cosmological epoch at the end
of inflation or even later which is the time when the cosmic
strings are produced and observed. In the proposal for heterotic
M-theory inflation made in \cite{Becker:2005sg}\footnote{Another
proposal has been made in \cite{Buchbinder:2004nt} and cosmic
strings were argued to arise from open membranes which stretch
between the visible M9 and an M5-brane just $10^{-4}L$ away from
the visible M9 \cite{Buchbinder:2005jy}.} which we will use here
the inflationary dynamics relies on the interactions between
several \MT branes in the bulk. Towards the end of inflation the
11-dimensional bulk gets however cleared of its \MT branes which
coalesce with the boundary M9's. This justifies the neglect of the
\MT branes in the flux background. Let us also note that the
addition of \MT branes would weaken the tight relation between the
GUT and gravity sector which relates so successfully the standard
values for $M_{GUT}\simeq 3\times 10^{16}$\,GeV and
$\alpha_{GUT}\simeq 1/25$ to the observed value for Newton's
Constant $G_N$.

We can now explicitly integrate over $x^{11}$ with the result that
the M2 brane action becomes the cosmic string action
\beqa
S_{M2} = \mu_{M2} \int_{\mathbf{R}^1} \!\!\! dt
\int_{\mathbf{S}^1} \!\!\! dx \sqrt{-g_{tt} g_{xx}}
+ \hdots \; .
\eeqa
with tension determined by the warp-factor and length $L$ of the
$\Orb$ interval
\beqa
\mu_{M2} = \tau_{M2} \int_0^L dx^{11} e^{-f(x^{11})/2}
= \frac{3\tau_{M2}}{2Q_v} \Big(1-(1-L Q_v)^{2/3}\Big) \; .
\label{M2tension}
\eeqa
To evaluate the value, let us remind that the correct value of the
4d Newton's Constant requires $L$ to be of critical length $L_c$
which is given in terms of the M9 charge by \cite{Curio:2000dw},
\cite{Curio:2003ur}
\beqa
L_c\equiv 1/Q_v \; .
\eeqa
We should therefore use $L\simeq L_c$ for the evaluation of the
cosmic string's tension. To evaluate the tension, let us express
all quantities in terms of the 11-dimensional gravitational
coupling constant $\kappa_{11}$. Based on phenomenological
reasoning the critical length will be given by
\cite{Banks:1996ss}, \cite{Curio:2003ur}
\beqa
L_c\simeq 12\kappa_{11}^{2/9} \; .
\label{CritLength}
\eeqa
With the M2 brane tension $\tau_{M2} = M_{11}^3/(2\pi)^2$, and the
defining relation $2\kappa_{11}^2 = (2\pi)^8/M_{11}^9$ for the 11d
Planck-mass $M_{11}$, we obtain for the string's tension
\beqa
\mu_{M2}
= \frac{3\tau_{M2}}{2 Q_v}
= 3L_c \Big(\frac{\pi}{2\kappa_{11}}\Big)^{\frac{2}{3}}
\simeq \, 9(2^{10}\pi^2)^{1/3} M_{GUT}^2 \; .
\eeqa
For the last expression we have used the relations
(\ref{GUTScale}) and (\ref{CritLength}). Since $\mu_{M2}^{1/2}$
turns out to be larger than the GUT scale, it is clear that the
string's tension comes out too large. This becomes evident when we
finally evaluate
\beqa
G_N\mu_{M2} \simeq 1.2 \times 10^{-3}
\eeqa
with $M_{GUT}\simeq 3\times 10^{16}$\,GeV which is in clear
conflict with the observational bound (\ref{Bound}). Also
considering a slightly smaller length $L = 11\kappa_{11}^{2/9}
= L_c-\kappa_{11}^{2/9}$, which could still be stabilized at the
end of inflation, would only decrease the tension by a factor of
$0.8$ which is not enough. The $\text{M2}_\perp$ candidates are
therefore ruled out as viable cosmic strings.

\subsection{$\text{M5}_\parallel$ Brane Case}

Let us now turn to the $\text{M5}_\parallel$ cosmic strings. The
Nambu-Goto term of the $\text{M5}_\parallel$ brane action reads
\beqa
S_{M5_\parallel} = \tau_{M5} \int_{\mathbf{R}^1} dt
\int_{\mathbf{R}^1} dx \int_{\mathbf{\Sigma}_4} d^4y \sqrt{-\det
h_{ab}}
\eeqa
where $a,b,\hdots = t,x,y^1,y^2,y^3,y^4$. Adopting again static
gauge for its embedding, we have to integrate over the 4-cycle
$\mathbf{\Sigma}_4$ to obtain the action for the cosmic string
\beqa
S_{M5_\parallel} = \mu_{M5,\parallel} \int_{\mathbf{R}^1} dt
\int_{\mathbf{R}^1} dx \sqrt{-g_{tt} g_{xx}}
\eeqa
with string tension given by
\begin{alignat}{3}
\mu_{M5_\parallel} &= \tau_{M5}e^{f(x^{11}_{M5})}
\int_{\mathbf{\Sigma}_4} d^4 y \, \Big(\!\!\prod_{i=1,\hdots,4}
g(X)_{y^i y^i}\Big)^{1/2}
\notag \\
&= \tau_{M5} \bigg(1-\frac{x_{M5}^{11}}{L_c}\bigg)^{\frac{2}{3}}
V_{\mathbf{\Sigma}_4}\;.
\end{alignat}
Here $0\le x_{M5}^{11}\le L$ denotes the position of the
$\text{M5}_\parallel$ along the $\Orb$ orbifold. It will be
convenient to write the volume of the 4-cycle
$V_{\mathbf{\Sigma}_4}$ in terms of a dimensionless radius
$r_{\mathbf{\Sigma}_4}$ by rescaling with the radius $R_v$ of $X$
on the visible boundary, i.e.~the undeformed initial Calabi-Yau
radius
\beqa
V_{\mathbf{\Sigma}_4} = \big(r_{\mathbf{\Sigma}_4} R_v
\big)^4\; .
\eeqa
Typically one would expect for a more or less isotropic Calabi-Yau
that $r_{\mathbf{\Sigma}_4} \lesssim 1$. For highly
anisotropic compactification spaces it could be larger.

To evaluate the tension's value, we need to employ another
standard relation \cite{Banks:1996ss}, \cite{Curio:2003ur}
\beqa
R_v \equiv V_v^{1/6} = 1/M_{GUT} \; .
\eeqa
Using this, the definition of the $\text{M5}_\parallel$ brane's
tension, $\tau_{M5} = M_{11}^6/(2\pi)^5$, plus (\ref{GUTScale}) we
arrive at
\beqa
\mu_{M5_\parallel} = 64 \Big(\frac{\pi}{2}\Big)^{\frac{1}{3}}
\bigg(1-\frac{x_{M5}^{11}}{L_c}\bigg)^{\frac{2}{3}}
M^2_{GUT}r^4_{\mathbf{\Sigma}_4} \; .
\eeqa
Numerically this leads to the following result
\beqa
G_N\mu_{M5_\parallel} = 4.7 \times 10^{-4}
\bigg(1-\frac{x_{M5}^{11}}{L_c}\bigg)^{\frac{2}{3}}
r^4_{\mathbf{\Sigma}_4} \; .
\label{M5Tension}
\eeqa
We will subsequently see that the production of the
$\text{M5}_\parallel$ cosmic strings will happen towards the end
of inflation essentially on the hidden M9 when $L$ gets stabilized
near $L_c$ \cite{Curio:2001qi}, \cite{Becker:2004gw}. Taking
therefore, say, $x^{11}_{M5} = L\simeq 11\kappa^{2/9}_{11} = L_c -
\kappa^{2/9}_{11}$, we obtain $G_N\mu_{M5_\parallel} = 8.9 \times
10^{-5} r^4_{\mathbf{\Sigma}_4}$. A radius
$r_{\mathbf{\Sigma}_4}\le 0.22$ would then already be enough to satisfy the observational constraint. Hence the $\text{M5}_\parallel$ easily passes the tension constraint. The positioning of the $\text{M5}_\parallel$ brane on the hidden boundary is also supported by the fact that M5 branes can only wrap 4-cycles which carry no $G$-flux \cite{Duff:1996rs}. In general this is the case on either the visible or hidden M9 boundary where the $G^{(2,2,0)}$ flux vanishes as a direct consequence of the $\Z2$ symmetry of the background.

\subsection{$\text{M5}_\perp$ Brane Case}

Let us finally come to the $\text{M5}_\perp$ cosmic strings. We
start from the $\text{M5}_\perp$ brane action
\beqa
S_{M5_\perp} = \tau_{M5} \int_{\mathbf{R}^1} dt
\int_{\mathbf{R}^1} dx \int_0^L dx^{11}
\int_{\mathbf{\Sigma}_3(x^{11})} \!\!\!\!\!\!\!\!\! d^3y \;\;
\sqrt{-\det h_{ab}}
\eeqa
Integrating over the compact dimensions gives the cosmic string
action
\beqa
S_{M5_\perp} = \mu_{M5_\perp} \int_{\mathbf{R}^1} dt
\int_{\mathbf{R}^1} dx \sqrt{-g_{tt} g_{xx}}
\eeqa
with string tension
\beqa
\mu_{M5_\perp} = \frac{3}{5}\tau_{M5}
\Big(1-(1-L/L_c)^{5/3}\Big) L_c
V_{\mathbf{\Sigma}_3}\;.
\eeqa
Again it will be convenient to express the volume of the 3-cycle
$V_{\mathbf{\Sigma}_3}$ through a dimensionless radius
$r_{\mathbf{\Sigma}_3}$ defined by
\beqa
V_{\mathbf{\Sigma}_3} = \big(r_{\mathbf{\Sigma}_3} R_v
\big)^3 \; .
\eeqa
With the standard relations used earlier we arrive then at
\beqa
\mu_{M5_\perp} = \frac{72}{5}\Big(\frac{\pi}{2}\Big)^{1/3}
\Big(1-(1-L/L_c)^{5/3}\Big) M_{GUT}^2
r^3_{\mathbf{\Sigma}_3} \; ,
\eeqa
which gives the result
\beqa
G_N\mu_{M5_\perp} = 1.1 \times 10^{-4}
\Big(1-(1-L/L_c)^{5/3}\Big) r^3_{\mathbf{\Sigma}_3} \; .
\eeqa

Again for a value $L=11\kappa_{11}^{2/9}$, we obtain
$G_N\mu_{M5_\perp} = 1.1 \times 10^{-4}r^3_{\mathbf{\Sigma}_3}$.
Hence the observational constraint can be satisfied for
$r_{\mathbf{\Sigma}_3}\le 0.12$. This still seems a rather mild
constraint on the average radius of the 3-cycle
$\mathbf{\Sigma}_3$. We can therefore conclude that also the
$\text{M5}_\perp$ cosmic strings pass the tension test. We will
next analyze the stability of our two M5 brane candidates.

\section{Stability}

\subsection{Classical Stability}

Cosmic strings resulting from fundamental heterotic strings were
found in \cite{Witten:1985fp} to be unstable. The reason was that
these cosmic strings are axionic strings with $\mathbf{S}^1$
topology which bound domain walls. Due to the domain wall tension
which is proportional to the area they span, these axionic strings
will quickly shrink. Hence they cannot become macroscopically
large.

We will at first sight encounter the same instability for cosmic
strings resulting from wrapped M2 or M5 branes in heterotic
M-theory. This is because these branes are charged under the
3-form $C_3$ resp.~dual 6-form potential $C_6$, which when reduced
over the appropriate cycle which the brane wraps, becomes a 2-form
potential $C_{[2]}$ in four dimensions. Since the dual of this
2-form gives an axion $\phi$ via
\beqa
dC_{[2]} = \star_4 d\phi \; ,
\eeqa
it seems that cosmic strings created by wrapping M2 or M5 branes
cannot grow to cosmic size due to their coupling to the axion
$\phi$. To avoid this conclusion one needs to remove the massless
axion. We will see that this will only be possible for the
$\text{M5}_\parallel$ cosmic string candidate and requires it to
be on the hidden M9. Hence the $\text{M5}_\perp$ cosmic string
candidate will be ruled out as it suffers from the domain wall
instability and therefore quickly shrinks to microscopic size. Let
us now explain how and under which conditions the massless axion
gets removed.

For this, let us remind first that the presence of the boundaries
in heterotic M-theory lead to a modification of its 4-form
field-strength $G$ on the boundaries \cite{Horava:1996ma}. This
modification involves the Yang-Mills and Lorentz Chern-Simons
3-forms $\omega_Y, \omega_L$ and one finds on the hidden
boundary\footnote{Since we will find that cosmic string production
will preferably occur close to the hidden boundary, we will focus
on this boundary here.} at $x^{11}=L$
\beqa
G_4 = dC_3 +
c\kappa^{2/3}_{11}\Big(\omega_Y-\frac{1}{2}\omega_L\Big)
\delta(x^{11}-L)\we dx^{11} \; , \qquad
c=\frac{\sqrt{2}}{(4\pi)^{5/3}}
\eeqa
To avoid carrying around the delta-function, let us write this in
10d notation in terms of the Neveu-Schwarz 3-form field-strength
$H$ on the hidden boundary (where $H_{ABC} = G_{11ABC}$, $B_{AB}
= C_{11AB}$)
\beqa
H_3 = dB_2-\frac{c\kappa_{11}^{2/3}}{2L}
\Big(\omega_Y-\frac{1}{2}\omega_L\Big) \;.
\eeqa
Since $\alpha'=2c\kappa_{11}^{2/3}/L$ \cite{Banks:1996ss}, we
recognize the familiar $\alpha'$ correction of the weakly coupled
heterotic string, with the difference of the factor $1/2$ which
arises from the separation of the boundaries. Plugging this
field-strength into the hidden boundary $\cM^{10}_h$ kinetic term
\beqa
-\frac{L}{2\kappa_{11}^2} \int_{\cM^{10}_{h}} H_3\we\star_{10} H_3
\eeqa
leads upon dualization $dC_6=\star_{10}dB_2$ to the coupling
\beqa
\frac{c}{2\kappa_{11}^{4/3}} \int_{\cM^{10}_h} C_6\we\Big(\tr F\we
F - \frac{1}{2}\tr R\we R\Big) \; .
\label{10Coupling}
\eeqa

We know that in order to stabilize the hidden boundary close to
the phenomenologically relevant length $L_c$ after inflation, the
hidden $E_8$ gauge symmetry must be broken to a gauge group of
smaller rank \cite{Becker:2004gw}. This will typically provide us
with some $U(1)$ gauge symmetries on the hidden M9. Let's pick one
of these and denote its field-strength $\cF_2=dA_1$. Moreover, let
us assume a non-vanishing gauge flux $\int_{{\cal C}_2}F\ne 0$
over some 2-cycle on $X$. Let us consider the coupling term
together with the kinetic terms in the 11-dimensional action
\beqa
\!\!\!\!\!\!-\frac{1}{2\times 7!\kappa_{11}^2}\int_{\cM^{11}}
|dC_6|^2
+ \frac{c}{2\kappa_{11}^{4/3}} \int_{\cM^{10}_h} C_6\we\Big(\tr
F\we F - \frac{1}{2}\tr R\we R\Big)
- \frac{1}{4g_{10}^2}\int_{\cM^{10}_h}|F|^2
\label{11action}
\eeqa
Here the 10-dimensional gauge coupling $g_{10}$ is fixed in terms
of the gravitational coupling as $g_{10}^2=(2^7\pi^5)^{1/3}
\kappa_{11}^{4/3}$ \cite{Horava:1996ma}. After a reduction to four
dimensions these terms will give a contribution (we will not
consider the curvature term $\tr R\we R$ further)
\beqa
-\frac{1}{2}\int_{\cM^4} |dC_{[2]}|^2
+ m \int_{\cM^4} C_{[2]} \we \cF_2
- \frac{1}{2}\int_{\cM^4}|\cF_2|^2
\label{4action}
\eeqa
to the 4-dimensional action. The mass parameter $m$ is given by
\beqa
m = \frac{(7!)^{1/2}}{2^{8/3}\pi^{5/6}} \times
\frac{\kappa_{11}^{1/3} L_{top}^4}{(L\langle V\rangle V_h)^{1/2}}
\; ,
\eeqa
where $\langle V\rangle$ denotes the Calabi-Yau volume averaged
over the $\Orb$ interval, $V_h$ represents the Calabi-Yau volume
at the location of the hidden boundary and the length $L_{top}$
will be defined next. To arrive at this expression, we have set
\begin{alignat}{3}
\int_{\cM_h^{10}} C_6\we\tr (\cF_2\we F) = L_{top}^4\int_{\cM^4}
C_{[2]}\we\cF_2
\end{alignat}
and then rescaled
\begin{alignat}{3}
\cF_2 &\rightarrow \Big(\frac{V_h}{2g_{10}^2}\Big)^{1/2} \cF_2 \\
C_{[2]} &\rightarrow
\Big(\frac{2\langle V\rangle L}{7!\kappa_{11}^2}\Big)^{1/2} C_{[2]}
\end{alignat}
such that the 4-dimensional fields $C_{[2]}$, $A_1$ receive a
canonical mass dimension one. The volume and length factors which
enter the rescaling originate from the ordinary reduction of the
metric dependent kinetic terms for $C_6$ and $A_1$ from 11
resp.~10 to 4 dimensions. The length parameter $L_{top}$ which
stems from the reduction of the metric independent topological
coupling term characterizes the localization of the gauge flux $F$
and $C_6$ on $X$.

It is now straightforward to demonstrate\footnote{Although we are
considering here the M5-M9 system, the following argumentation
closely parallels the argumentation for the D1-D3 brane case
\cite{Copeland:2003bj}, \cite{Leblond:2004uc}.} that this action
implies the absence of the axion $\phi$ which we will show next.
The field equations for $A_1$ and $C_{[2]}$ which result from the
action (\ref{4action}) are
\begin{alignat}{3}
d\star_4 dA_1 &= -m dC_{[2]}\\
d\star_4 dC_{[2]} &= -m\cF_2 \; .
\end{alignat}
We can solve the second equation by
\beqa
dC_{[2]} = \star_4(d\phi - m A_1) \; ,
\label{Sol}
\eeqa
which defines the dual axion field $\phi$. Plugging this solution
back into the field equation for $A_1$ gives
\beqa
d\star_4 dA_1 = \star_4 (- m d\phi + m^2 A_1) \; .
\eeqa
For the ground state in which $\phi=0$ or by picking a gauge which
sets $d\phi=0$, this result shows that $A_1$ has acquired a mass
$m$. Alternatively, one might plug the solution back into the
action (\ref{4action}). Then the coupling term gives us a mass
term for $A_1$
\beqa
m\int_{\cM^4} C_{[2]}\we dA_1
= \int_{\cM^4} (m A_1\we \star_4 d\phi - m^2 A_1\we\star_4 A_1)
\; .
\eeqa
Furthermore, we infer from (\ref{Sol}) that $\phi$ must transform
nonlinearly under $A_1$ gauge transformations
\beqa
\delta A_1=d\Lambda\; , \qquad \delta\phi=-m\Lambda \; .
\eeqa

The proper interpretation of these results is that the $U(1)$
gauge field swallows the axion $\phi$, gains a further degree of
freedom and becomes massive, i.e.~$A_1\rightarrow A_1-d\phi/m$.
Since the axion gets removed in this Higgsing, there is no domain
wall anymore which would prevent the cosmic string from growing.
Let us note that $m$ grows when the hidden boundary comes close to
the critical length $L_c$ where $V_h$ would classically vanish and
quantum-mechanically is expected to reach Planck-size\footnote{The
11-dimensional Planck-length $l_{11}$ is defined by
$2\kappa_{11}^2 = 16\pi G_{N,11} = (2\pi)^8 l_{11}^9$.}
$l_{11}^6\simeq \big(\frac{\kappa_{11}^{2/9}}{5}\big)^6$. Since
towards the end of the inflationary mechanism of
\cite{Becker:2005sg} the hidden boundary gets indeed stabilized
close to $L_c$, where $V_h$ becomes small, through the
stabilization mechanisms developed in \cite{Curio:2001qi},
\cite{Becker:2004gw} we notice that the removal of the axion
domain wall will be particularly effective towards the end of
inflation when $m$ becomes large.

For which of our cosmic string candidates, $\text{M5}_\parallel$, $\text{M5}_\perp$ does this stabilization mechanism apply? The gauge fields $F$ are localized on the boundary and therefore the initial coupling (\ref{10Coupling}) will only be non-vanishing for a parallel $\text{M5}_\parallel$ brane which moreover has to be localized on the hidden boundary. The transverse $\text{M5}_\perp$ which stretches orthogonal to $\cM_h^{10}$ along $\Orb$ cannot have this coupling. It will therefore maintain its domain wall instability and will consequently quickly shrink to microscopic size. This might have been anticipated because the $\text{M5}_\perp$ is a non-BPS object in flat 11-dimensional spacetime. We are therefore left with a unique cosmic string candidate, a parallel $\text{M5}_\parallel$ brane on the hidden boundary.

Let us now come to a second potential instability which is the
breaking of the $\text{M5}_\parallel$ cosmic string on the hidden
boundary. Since the endpoints which are produced when the string
breaks are still connected by flux lines, one can think of this
breaking as the $\text{M5}_\parallel$ brane dissolving in the M9.
One has to compare the gauge flux $\int_{{\cal C}_2}F$ which is
transverse to the $\text{M5}_\parallel$ brane with the kinetic
energy density $\int_X F\we\star_6 F$ on $X$. By counting
dimensions one would conclude that it might be energetically
favorable for the flux to expand along $X$ and therefore the
cosmic string might break.

The reason why this conclusion should not hold is very simple.
Notice that the argument so far implicitly assumed that $X$ is
large enough in order to provide space for the flux to spread
along $X$. This, however, is not the case precisely on the hidden
M9. As we will review later, $L$, and therefore the hidden M9,
gets stabilized towards the end of inflation close to $L_c$. The
characteristic feature of $L_c$ is that it is the length at which
the volume of $X$ shrinks classically to a point. Therefore the
flux has no space to spread along $X$ when the
$\text{M5}_\parallel$ brane is on (or close to) the hidden M9.
Another argument against the breaking of the string, even at
finite size $X$ volumes, might also come from the nice solution of
the breaking instability for a D1 on a D3 brane presented in
\cite{Leblond:2004uc}. Here, as well as in our case we have a flux
$\int_{{\cal C}_2} F\ne 0$ transverse to the cosmic resp.~D1
string. Since we have, however, not a volume for $X$ of sizeable
size, we will not explore this possibility further here.

So it remains to analyze whether there can be breakage of the
$\text{M5}_\parallel$ cosmic string in the four non-compact
directions. Here let us note that the $\text{M5}_\parallel$ cosmic
strings, when being located on the hidden M9, lead in four
dimensions to an effective abelian Higgs model whose $U(1)$ is
Higgsed. Consequently, Abrikosov-Nielsen-Olesen type flux tubes \cite{Abrikosov:1956sx} will form which carry magnetic flux of the Higgsed $U(1)$. These flux tubes, in which the field strength falls off exponentially with radial distance, cannot decay because they are topologically stable. It is these flux tubes which represent the $\text{M5}_\parallel$ cosmic strings in the effective
four-dimensional theory and show that they are also stable with respect to breakage along the non-compact directions. One might worry that at high energies when the gauge theory on the hidden M9 is expected to restore a GUT symmetry\footnote{Notice that at low energies the hidden M9 does not carry a GUT theory since the resulting stabilized orbifold length would be too short and the supersymmetry breaking scale much larger than TeV  \cite{Becker:2004gw}.} with a corresponding embedding of the $U(1)$ into the unified gauge group, the flux tubes might break. The reason being that GUT theories possess monopoles such that the flux tube can start on a monopole and end on an anti-monopole, thus making it unstable against monopole pair production. An estimate of the monopole pair creation rate via the Schwinger pair production calculation shows, however, that this rate is suppressed by a factor $\text{exp}\big(-\frac{\pi M^2} {\mu_{\text{M5}_\parallel}}\big)$ with $M$ being the monopole mass. We expect the $\text{M5}_\parallel$ cosmic string's tension $\mu_{\text{M5}_\parallel}$ to be far smaller than the monopole's
mass, again due to its warp-factor suppression. Therefore the
scale of the monopole mass should easily be an order of magnitude
larger than the scale of the string's tension which is enough to
render the flux tubes effectively stable on cosmological time
scales \cite{Polchinski:2004ia}. Before describing how the
parallel $\text{M5}_\parallel$ branes are produced when inflation
comes to an end, we will now briefly address the stability of M2
branes and quantum instabilities.

Though we have seen that the tension of an M2 cosmic string violates the observational bound and M2 cosmic strings are consequently ruled out, let us as nevertheless include the stability discussion for a hypothetical M2 cosmic string. In this case there is a similar coupling, the well-known \cite{Horava:1996ma}
\beqa
\frac{\sqrt{2}}{(4\pi)^3(4\pi\kappa_{11}^2)^{1/3}}\int_{\cM^{11}}
C_3\we X_8(F,R)
\eeqa
where
\beqa
\qquad X_8(F,R) = -\frac{1}{4}\Big(\tr F^2 -\frac{1}{2}\tr R^2\Big)^2 + \Big(-\frac{1}{8}\tr R^4 +\frac{1}{32}(\tr R^2)^2\Big) \eeqa
Combining it with the kinetic terms for $C_3$ and $F$ can once again generate the desired effective 4d coupling $\int_{\cM^4}C_{[2]}\we \cF_2$. This time it requires an orthogonal $\text{M2}_\perp$ brane because $F$ is localized on the boundary M9's. Assuming a non-zero higher instanton charge $\int_X(F\we F\we F) \ne 0$ on $X$ we would likewise remove the axion and the associated domain wall through Higgsing of the 4-dimensional $U(1)$. This time we have a topological charge $\int_X (F\we F\we F)$ on $X$ which we need to compare to the energy density term $\int_X F\we\star_6 F$. Counting dimensions, we would conclude that it is energetically favorable for the flux to shrink. Hence, the hypothetical M2 cosmic string would not break up as it cannot transform into flux which can spread out over the M9. We will later also see that transverse branes will not be produced at the end of inflation. The stability of the $\text{M2}_\perp$ brane will therefore not imply its presence.

\subsection{Quantum Stability}

One might ask whether the $\text{M5}_\parallel$ cosmic strings
could decay quantum-mechanically via some non-perturbative effect.
With only M2 and M5 brane instantons available, this would require
that either of them must be able to couple to the
$\text{M5}_\parallel$ brane. For the M2 instantons\footnote{See
\cite{Carlevaro:2005bk} for a recent discussion of these
instantons.} to mediate a force, they would need to wrap a genus
zero holomorphic 2-cycle {\STN} on the divisor {\SF}. Hence, if
the divisor {\SF} does not contain any such 2-cycles
{\STN}, the $\text{M5}_\parallel$ brane and thus the cosmic string
would not feel a force mediated by M2 instantons. Moreover, no M5
instantons i.e.~M5 branes which wrap the complete $X$ at some fixed
location along the $\Orb$ can attach to the $\text{M5}_\parallel$
branes because the M5 instantons would need two more compact
dimensions than the divisor which the $\text{M5}_\parallel$ wraps
can provide. Consequently, M5 instantons will not be able to exert a force on the $\text{M5}_\parallel$ branes. Therefore with respect to M2 or M5 instanton decay the $\text{M5}_\parallel$ cosmic strings are stable as long as the divisor {\SF} does not contain any genus zero holomorphic 2-cycles {\STN}.

\subsection{Relation to Other Types of Cosmic Strings}

Cosmic D-strings which arise from the tachyon condensation of a brane-antibrane Dp-$\bar{\text{D}}$p pair have a priori a very different fundamental description from the heterotic cosmic strings originating from wrapped M5 branes. At the level of the effective 4-dimensional description there are, however, striking similarities. Let us consider for definiteness a D3-$\bar{\text{D}}$3 pair on whose worldvolume a D1-string forms as a tachyonic vortex \cite{Sen:1998ii}. The tachyon in the open string spectrum of the D3-$\bar{\text{D}}$3 system is charged under the diagonal combination of the two $U(1)$'s. When the tachyon condenses in a topologically non-trivial vacuum the diagonal $U(1)$ is Higgsed. The effective picture \cite{Blanco-Pillado:2005xx} of the created D1-string is a topologically stable vortex solution which carries magnetic flux of the Higgsed $U(1)$ similar to an Abrikosov-Nielsen-Olesen flux tube \cite{Abrikosov:1956sx}. The Ramond-Ramond charge of the D1-string stems from a Wess-Zumino coupling
\beqa
\int_{D3-\bar{\text{D}}3} \cF_2 \we C_2
\label{4Coupling}
\eeqa
on the D3-$\bar{\text{D}}$3 worldvolume. Here, $\cF_2$ denotes the field-strength of the diagonal $U(1)$ and $C_2$ the Ramond-Ramond 2-form. In four dimensions the D1-string represents a cosmic string \cite{Copeland:2003bj}. Hence, together with the kinetic terms for the gauge potential and $C_2$ we arrive at an effective action which is formally the same as in (\ref{4action}). Consequently, both the heterotic cosmic strings and the type II cosmic D-strings have the same effective description in terms of Abrikosov-Nielsen-Olesen type flux tubes. Indeed the analogy between both can be extended further as we will now indicate.

Solitonic descriptions of cosmic superstrings had been given in \cite{Binetruy:1998mn}, \cite{Davis:2005jf} for heterotic string motivated models and in \cite{Dvali:2002fi}, \cite{Dvali:2003zh},
\cite{Blanco-Pillado:2005xx}, \cite{Achucarro:2005vz} for
D-strings. Although the low-energy effective actions are very similar in both cases, they differ by a dilaton-independent D-term contribution from a Fayet-Iliopoulos term $\xi$ of the Higgsed $U(1)$. This Fayet-Iliopoulos term $\xi$ was not obvious and therefore omitted in the heterotic models \cite{Binetruy:1998mn}, \cite{Davis:2005jf} while it was included for the type II D1-string, being proportional to the D3-brane tension \cite{Blanco-Pillado:2005xx}. The presence of this term is crucial as it allows to construct solitonic supersymmetric solutions free of singularities \cite{Blanco-Pillado:2005xx}. With the construction of heterotic cosmic strings in terms of wrapped M5 branes, it is natural to guess that the $\text{M5}_\parallel$ tension could provide this Fayet-Iliopoulos term on the heterotic side. Furthermore, one might wonder whether the effective heterotic M-theory action (\ref{4action}) could be extended to include a tachyon like in the effective D3-$\bar{\text{D}}$3 or D1-D3 descriptions with the tachyon playing the role of the Higgs field. This seems indeed the case. Similar to the type II D3-$\bar{\text{D}}$3 or D1-D3 systems where the tachyon appears when both branes are close to each other, there are fields $\Phi$ in heterotic M-theory coming from M2 branes stretching between the $\text{M5}_\parallel$ brane and the hidden M9. These fields acquire a negative mass squared and hence indeed become tachyonic when the  $\text{M5}_\parallel$ brane comes close to the M9 \cite{Buchbinder:2004nt}.

It might also be interesting to study whether viable cosmic strings originating from wrapped M5-branes may also arise in M-theory compactifications on $G_2$ manifolds. We will mention just a few aspects and leave a full investigation to future work. First, in contrast to the heterotic M-theory case, $G_2$ compactifications preserving an $N=1$ supersymmetry must have zero $G$-flux and hence possess no warping \cite{deWit:1986xg}, \cite{Acharya:2000ps}. The smallness of the cosmic string tension must therefore arise from a combination of a low (as compared to the 4-dimensional Planck-scale) fundamental scale $1/\kappa_{11}^{2/9}$ together with the presence of a 4-cycle of sufficiently small volume. Indeed for special cases \cite{Friedmann:2002ty} a low fundamental scale $1/\kappa_{11}^{2/9}$ close to the GUT scale has been confirmed. Second, phenomenologically viable $G_2$ compactifications with non-abelian gauge groups of type $A,D$ or $E$ and charged chiral matter require the presence of a 3-dimensional locus $Q$ of $A,D$ or $E$ orbifold singularities on the $G_2$ manifold. $Q$ itself is smooth but the normal directions to $Q$ have a singularity. It remains, however, an open problem \cite{Acharya:2004qe} to construct compact $G_2$ manifolds with such singularities. Consequently, the full effective 4-dimensional theory is not known to date. Anomaly considerations \cite{Witten:2001uq} reveal in the case of $A_n=SU(n+1)$ gauge groups a 7-dimensional interaction term
\beqa
\int_{\cM^4\times Q} K\wedge \Omega_5(A)
\eeqa
with $K$ the 2-form field strength of a $U(1)$ gauge field which is part of the normal bundle to $Q$ and $\Omega_5(A)$ the Chern-Simons 5-form satisfying $d\Omega_5(A) = \tr F\we F \we F$. This term does not lead, in contrast to the heterotic M-theory case with Green-Schwarz anomaly cancelling terms, to a coupling of type (\ref{4Coupling}) needed to gauge away the axion and therefore the domain wall instability of the M5 brane cosmic string. The stability of M5 brane wrapped cosmic strings is therefore not clear in M-theory on $G_2$ manifolds. One should also add that a viable model of inflation arising from such M-theory compactifications has still to be constructed.

\section{End of Inflation}

So far we have systematically analyzed which cosmic string
candidates pass the tension constraint and the stability
criterion. The only candidate left over is a parallel
$\text{M5}_\parallel$ brane localized on the hidden boundary. It
remains to clarify whether these branes can also be produced
towards the end of inflation. Let us therefore now briefly provide
some background on the end of heterotic M-theory inflation
following \cite{Becker:2005sg}.

The inflationary phase is driven through non-perturbative
interactions between several \MT branes distributed along the
$\Orb$ interval. Initially close together, the repulsive
interactions between neighboring \MT branes drags them towards the
boundaries. This characterizes the inflationary phase. The fact
that many \MT branes are present enhances the Hubble friction and
leads to an M-theory realization of the assisted inflation idea
\cite{Liddle:1998jc} with parametrically small slow-roll
parameters. As long as the distance between the \MT branes stays
smaller than the orbifold length $L$, the resulting potential
assumes the required simple exponential form.

This changes at the end of inflation. Here the distances between
the \MT branes have grown to a size comparable to that of the
$\Orb$ length $L$ itself and further contributions to the dynamics
of the system become equally relevant. These are, a repulsive open
M2 instanton force mediated between the two boundary M9's, gaugino
condensation and fluxes. Let us detail this a bit more. The fact
that at this stage the repulsive M9-M9 interaction becomes
noticeable causes $L$ to grow. Characteristic for heterotic
M-theory, a growing $L$ implies a growing gauge coupling on the
hidden M9. This is a consequence of the theory's warped flux
compactification background \cite{Curio:2000dw},
\cite{Curio:2003ur}, \cite{Krause:2001qf}, \cite{Witten:1996mz}.
Hence towards the end of inflation the hidden gauge theory becomes
strongly coupled which triggers gaugino condensation. As a
consequence of gaugino condensation a non-vanishing Neveu-Schwarz
H-flux will be induced on the hidden M9. This is due to a specific
perfect square structure within the heterotic action which
combines gaugino condensation and $H$-flux \cite{Horava:1996vs}
(recent discussions can also be found in \cite{Gukov:2003cy},
\cite{Brustein:2004xn}, \cite{Curio:2005ew}, \cite{Frey:2005zz}).

The great importance of these additional contributions to the
potential which enter the stage only at the end of inflation --
M9-M9 interaction, gaugino condensation and $H$-flux -- lies in
the fact that they will stabilize the $\Orb$ length (``dilaton'')
and the Calabi-Yau volume (see e.g.~\cite{Gukov:2003cy},
\cite{Brustein:2004xn}, \cite{Curio:2005ew}, \cite{Frey:2005zz},
\cite{Becker:2003yv}, \cite{Buchbinder:2003pi},
\cite{Strominger:1986uh}, \cite{Becker:2005nb}, \cite{Curio:2006dc}). Most relevant for
us will be the $\Orb$ length $L$. Furthermore, in vacua with
positive vacuum energy $L$ will be stabilized close to its
critical length $L_c$ which is the length at which the hidden
Calabi-Yau volume vanishes classically. This can be achieved
either with help of one remaining position-stabilized \MT brane in
the bulk \cite{Curio:2001qi} or by breaking the $E_8$ gauge
symmetry on the hidden boundary \cite{Becker:2004gw}\footnote{The
resulting de Sitter vacua have been shown to be stable under
higher order $R^4$ corrections which amount to changes of a small
percentage \cite{Anguelova:2005jr}.}. A stabilization close to
$L_c$ is actually necessary to obtain a realistic value for
Newton's Constant and a supersymmetry breaking scale close to the
TeV scale \cite{Becker:2004gw}. The stabilization of $L$ close to
$L_c = 12\kappa_{11}^{2/9}$, say in a regime
\beqa
L_c - \kappa_{11}^{2/9} \le L \le L_c
\eeqa
has, however, an immediate impact on the cosmic string tensions
derived earlier. Let's focus on the viable $\text{M5}_\parallel$
cosmic string where $x^{11}_{M5}=L$ because we have seen that only
on the boundary\footnote{Since the tension will be lower on the
hidden than on the visible boundary, we take the hidden M9. The
fact that an $x^{11}_{M5}$ dependence arises in the tension might
not be surprising given that the $\text{M5}_\parallel$ brane
breaks the $N=1$ supersymmetry in four dimensions.} can it be freed
of its domain wall instability. From (\ref{M5Tension}) we see that
for $L\rightarrow L_c$ these cosmic strings can become nearly
tensionless. Such a low tension is only possible through the
warp-factor of the background which contributes the
$(1-x^{11}_{M5}/L_c)^{2/3}$ suppression factor to
(\ref{M5Tension}).

Let us conclude this section by stressing the salient feature of
this quick review. Namely, we can influence the tension of the
$\text{M5}_\parallel$ cosmic string by the value at which $L$ will
be stabilized at the end of inflation. Realistic stabilizations require a stabilization close to $L_c$ which lowers the cosmic string's tension considerably.

\section{Production}

Earlier we found that $\text{M2}_\perp$ cosmic strings would violate the observational bound on the cosmic string's tension. It is therefore satisfying to see that they are not being produced when inflation comes to an end. This is due to the fact that their
production would exceed the energy threshold being available at this time which lies certainly below $M_{GUT}$. We had further seen that the tension of an $\text{M5}_\parallel$ brane is small enough so that they can reach cosmic size once they are produced. In this
section we will qualitatively describe a mechanism which leads to
the production of these heterotic cosmic strings.

The model of inflation of \cite{Kachru:2003sx} is based on the
dynamics of a pair of D3 and anti-D3 branes. Towards the end of
inflation the distance between the brane and the anti-brane goes to
zero resulting in their annihilation. It has been argued in
\cite{Copeland:2003bj} that this annihilation results in the
creation of D1-branes which can reach a cosmic size.

The mechanism leading to cosmic string production in our scenario
is rather different and is based on the strongly time dependent
background which originates at the end of the inflationary process
\cite{Gubser:2003hm}, \cite{Gubser:2003vk}, \cite{Chung:1998bt},
\cite{Lawrence:1995ct}. The heterotic M-theory inflation model
presented in \cite{Becker:2005sg} is based on the dynamics of a
set of M5$_2$-branes which towards the end of inflation approach
the boundaries of the $\Orb$ interval. When the M5$_2$-branes hit
the boundaries the background becomes strongly time dependent and
at this point the inflaton field starts performing rapid coherent
oscillations with a Planck sized amplitude. Precisely these
oscillations provide the source of energy to pair produce strings
of low tension. The production rate for these strings was
evaluated in \cite{Gubser:2003hm}, \cite{Gubser:2003vk} from the
physical state constraint for the string states
\beqa
L_0\, |\text{physical}\rangle = 0,
\eeqa
which was rewritten as a differential equation for a string state
$\chi(t)$ who's oscillation frequency $\omega(t)$ is sourced by the
inflaton
\beqa
\ddot \chi+\omega (t)\chi=0.
\eeqa
It turns out that the pair produced strings cannot be fundamental
strings as their tension would be of the order of the
four-dimensional Planck scale, roughly $M_{Pl}\simeq 10^{18}$ GeV,
several orders of magnitude above the typical inflaton mass
$m_{inf}\simeq 10^{13}$ GeV.

Non-perturbative strings would be the alternative and these are
precisely the objects produced at the end of our inflationary
process. Indeed, our candidates for cosmic strings are not
fundamental strings but branes wrapped on a four-cycle of the
Calabi-Yau manifold. Towards the end of inflation the volume of the
four-cycle becomes very small as the Calabi-Yau volume shrinks to
very small size, endowing the corresponding strings with a low
tension. There is an extensive production of this type of strings (a
similar situation for non-perturbative strings obtained by wrapping
$D3$ branes on shrinking two-cycles has been discussed in
\cite{Gubser:2003hm} and references therein). A very rough estimate
shows that the effective tension of a string obtained by wrapping a
brane on a non-trivial cycle has to satisfy
\begin{equation}\label{ai}
\sqrt{\mu_{string}} \leq {1 \over 20} M_{Pl},
\end{equation}
in order to lead to a massive string production. This bound can be
easily satisfied for the case of an $\text{M5}_\parallel$ brane. In
this case the effective string tension is given by
\beqa
\mu_{\text{M5}_\parallel}
= \tau_{M5}(1-x_{M5}^{11}/L_c)^{\frac{2}{3}} V_{\mathbf{\Sigma}_4}\; .
\eeqa
This expression makes it clear that we can easily satisfy the
bound (\ref{ai}) by being close enough to the hidden boundary
where the warp factor can be made arbitrarily small. As a result
tensionless cosmic strings will be produced on the hidden
boundary. Even though the tensionless strings are produced on the
hidden boundary they still would have an effect on our visible
universe since they interact gravitationally. These strings would
then represent an interesting new dark matter candidate (for their
detection via gravitational lensing see
e.g.~\cite{Huterer:2003ze}, \cite{Damour:2004kw},
\cite{Fairbairn:2005zs}) next to other potential dark matter
residing on the hidden boundary \cite{Krause:2004up}. One final
remark on the stability of the pair produced strings is in order.
For the pair produced strings to be observed, it is important that
they stay around long enough and do not annihilate shortly after
being pair produced. Even though annihilation and decay of strings
are still poorly understood one can make several arguments in
favor of the stability and observability of the strings being pair
produced. One qualitative argument is based on the dimensionality
of the string world-sheet and was used many years ago by
Brandenberger and Vafa to argue that our world is four-dimensional
\cite{Brandenberger:1988aj}. We could argue that the odds for an
infinitely extended string pair to meet once produced are pretty
small, as the world-sheet of a string is two-dimensional and two
strings would only collide at an instant. A generalization of this
idea, worked out more recently in a paper by Randall and Karch
\cite{Karch:2005yz}, would allow us to exclude the production of
higher dimensional branes, as the odds for such branes to meet and
annihilate are much higher. It would be interesting to work out
the details of this higher order annihilation process more
precisely. We hope to report on this and on the production rate
calculation of the cosmic string candidate presented in this paper
elsewhere.

\bigskip
\noindent {\large \bf Acknowledgements}\\[2ex]
We would like to thank G.~Curio, S.~Gubser, A.~Mazumdar, L.~McAllister, G.~Moore, R.~Myers, L.~Randall, T.~Vachaspati and A.~Vilenkin for helpful discussions and correspondence. We would especially like to thank J.~Polchinski for a careful reading of the manuscript and many helpful discussions. Many thanks to the Perimeter Institute for Theoretical Physics, the Aspen Center for Physics, the Physics Department at Harvard University and the Radcliffe Institute for Advanced Studies for hospitality during different stages of this work. The work of K.~Becker was supported by NSF grant PHY-0244722, an Alfred Sloan Fellowship, the University of Texas A\&M and the Radcliffe Institute for Advanced Study of Harvard University. The work of M.~Becker was supported by NSF grant PHY-0354401, an Alfred Sloan Fellowship, the University of Texas A\&M and the Radcliffe Institute for Advanced Study of Harvard University. The work of A.K.~was supported by the NSF under grant PHY-0354401 and the University of Texas A\&M.

 \newcommand{\zpc}[3]{{\sl Z. Phys.} {\bf C\,#1} (#2) #3}
 \newcommand{\npb}[3]{{\sl Nucl. Phys.} {\bf B\,#1} (#2) #3}
 \newcommand{\plb}[3]{{\sl Phys. Lett.} {\bf B\,#1} (#2) #3}
 \newcommand{\prd}[3]{{\sl Phys. Rev.} {\bf D\,#1} (#2) #3}
 \newcommand{\prb}[3]{{\sl Phys. Rev.} {\bf B\,#1} (#2) #3}
 \newcommand{\pr}[3]{{\sl Phys. Rev.} {\bf #1} (#2) #3}
 \newcommand{\prl}[3]{{\sl Phys. Rev. Lett.} {\bf #1} (#2) #3}
 \newcommand{\jhep}[3]{{\sl JHEP} {\bf #1} (#2) #3}
 \newcommand{\jcap}[3]{{\sl JCAP} {\bf #1} (#2) #3}
 \newcommand{\cqg}[3]{{\sl Class. Quant. Grav.} {\bf #1} (#2) #3}
 \newcommand{\prep}[3]{{\sl Phys. Rep.} {\bf #1} (#2) #3}
 \newcommand{\fp}[3]{{\sl Fortschr. Phys.} {\bf #1} (#2) #3}
 \newcommand{\nc}[3]{{\sl Nuovo Cimento} {\bf #1} (#2) #3}
 \newcommand{\nca}[3]{{\sl Nuovo Cimento} {\bf A\,#1} (#2) #3}
 \newcommand{\lnc}[3]{{\sl Lett. Nuovo Cimento} {\bf #1} (#2) #3}
 \newcommand{\ijmpa}[3]{{\sl Int. J. Mod. Phys.} {\bf A\,#1} (#2) #3}
 \newcommand{\rmp}[3]{{\sl Rev. Mod. Phys.} {\bf #1} (#2) #3}
 \newcommand{\ptp}[3]{{\sl Prog. Theor. Phys.} {\bf #1} (#2) #3}
 \newcommand{\sjnp}[3]{{\sl Sov. J. Nucl. Phys.} {\bf #1} (#2) #3}
 \newcommand{\sjpn}[3]{{\sl Sov. J. Particles \& Nuclei} {\bf #1} (#2) #3}
 \newcommand{\splir}[3]{{\sl Sov. Phys. Leb. Inst. Rep.} {\bf #1} (#2) #3}
 \newcommand{\tmf}[3]{{\sl Teor. Mat. Fiz.} {\bf #1} (#2) #3}
 \newcommand{\jcp}[3]{{\sl J. Comp. Phys.} {\bf #1} (#2) #3}
 \newcommand{\cpc}[3]{{\sl Comp. Phys. Commun.} {\bf #1} (#2) #3}
 \newcommand{\mpla}[3]{{\sl Mod. Phys. Lett.} {\bf A\,#1} (#2) #3}
 \newcommand{\cmp}[3]{{\sl Comm. Math. Phys.} {\bf #1} (#2) #3}
 \newcommand{\jmp}[3]{{\sl J. Math. Phys.} {\bf #1} (#2) #3}
 \newcommand{\pa}[3]{{\sl Physica} {\bf A\,#1} (#2) #3}
 \newcommand{\nim}[3]{{\sl Nucl. Instr. Meth.} {\bf #1} (#2) #3}
 \newcommand{\el}[3]{{\sl Europhysics Letters} {\bf #1} (#2) #3}
 \newcommand{\ap}[3]{{\sl Annals Phys.} {\bf #1} (#2) #3}
 \newcommand{\jetp}[3]{{\sl JETP} {\bf #1} (#2) #3}
 \newcommand{\jetpl}[3]{{\sl JETP Lett.} {\bf #1} (#2) #3}
 \newcommand{\acpp}[3]{{\sl Acta Physica Polonica} {\bf #1} (#2) #3}
 \newcommand{\sci}[3]{{\sl Science} {\bf #1} (#2) #3}
 \newcommand{\vj}[4]{{\sl #1~}{\bf #2} (#3) #4}
 \newcommand{\ej}[3]{{\bf #1} (#2) #3}
 \newcommand{\vjs}[2]{{\sl #1~}{\bf #2}}
 \newcommand{\hepph}[1]{{\sl hep--ph/}{#1}}
 \newcommand{\desy}[1]{{\sl DESY-Report~}{#1}}

\bibliographystyle{plain}

\end{document}